\documentclass[12pt]{article}
\usepackage{amsmath}
\usepackage{amsfonts}
\DeclareMathSymbol{\mg}{\mathrel}{symbols}{"1D}

\newcommand{\ga}{\alpha}
\newcommand{\gb}{\beta}
\renewcommand{\gg}{\gamma}
\newcommand{\gd}{\delta}
\renewcommand{\ge}{\epsilon}

\newcommand{\gc}{\chi}
\newcommand{\gx}{\xi}
\newcommand{\gm}{\mu}
\newcommand{\gn}{\nu}

\newcommand{\gl}{\lambda}
\newcommand{\gr}{\rho}
\newcommand{\gth}{\theta}
\newcommand{\gs}{\sigma}

\newcommand{\go}{\omega}
\newcommand{\gz}{\zeta}

\newcommand{\gps}{\psi}

\newcommand{\gG}{\Gamma}
\newcommand{\gD}{\Delta}
\newcommand{\gF}{\Phi}

\newcommand{\gL}{\Lambda}
\newcommand{\gS}{\Sigma}
\newcommand{\gTh}{\Theta}
\newcommand{\gO}{\Omega}

\newcommand{\cD}{{\cal D}}

\newcommand{\cF}{{\cal F}}
\newcommand{\cG}{{\cal G}}
\newcommand{\cH}{{\cal H}}

\newcommand{\cL}{{\cal L}}

\newcommand{\cZ}{{\cal Z}}
\newcommand{\ugb}{{\underline\beta}}
\newcommand{\ugg}{{\underline\gamma}}

\newcommand{\bz}{{\bar z}}

\newcommand{\bF}{{\bar F}}

\newcommand{\bR}{{\bar R}}

\newcommand{\bge}{{\bar\epsilon}}

\newcommand{\bgc}{{\bar\chi}}

\newcommand{\bgz}{{\bar\zeta}}

\newcommand{\bgps}{{\bar\psi}}

\newcommand{\bgG}{{\bar\Gamma}}
\newcommand{\bgD}{{\bar\Delta}}

\newcommand{\bgS}{{\bar\Sigma}}

\newcommand{\slashed}{\hspace{-1.1ex}/} 
\newcommand{\Slashed}{\hspace{-1.7ex}/\hspace{.6ex}}

\newcommand{\Id}{\text{\small 1}\hspace{-3.5pt}\text{1}}

\newcommand{\der}{\partial}

\newcommand{\Der}{D}
\newcommand{\sDer}{\Der\Slashed} 
\newcommand{\sder}{\der\slashed} 
\newcommand{\nit}{\noindent}

\newcommand{\np}{\newpage}

\newcommand{\bit}{\bibitem}
\newcommand{\lh}{\left(}
\newcommand{\rh}{\right)}

\newcommand{\vs}[1]{\vspace{#1 em}} 
\newcommand{\labl}[1]{\label{#1}}
\newcommand{\Kh}{K\"{a}hler}

\newcommand{\beq}{\begin{gather}}
\newcommand{\eeeeeq}{\end{gather}}
\newcommand{\barr}{\begin{array}}
\newcommand{\earr}{\end{array}}
\newcommand{\equ}[1]{\begin{gather} #1 \end{gather}}

\newcommand{\mtrx}[1]{\begin{matrix} #1 \end{matrix}}

\newcounter{oldcounter}

\newcommand{\be}{\begin{equation}}
\newcommand{\ee}{\end{equation}}
\newcommand{\bea}{\begin{eqnarray}}
\newcommand{\eea}{\end{eqnarray}}

\newcommand{\commentary}[1]{}

\begin{document} 

\pagestyle{empty} 

\begin{flushright}
NIKHEF/2003-009\\
%hep-th/0307284\\
%version \today
\end{flushright} 
\vs{3} 
\begin{center} 
{\Large{\bf Supersymmetric hydrodynamics}} \\
\vspace{3ex}
{\large T.S.\ Nyawelo$^\dag$}\\
\vspace{3ex}
NIKHEF, PO Box 41882 \\ 
1009 DB Amsterdam \\
The Netherlands \\
\vs{1} 
%\today 
\vs{2}

{\small{ \bf{Abstract} }} \\
\end{center}

\nit
{\footnotesize{We work out some properties of a recently 
proposed globally $N = 1$ supersymmetric extension of relativistic 
fluid mechanics in four-dimensional Minkowski space. We construct the 
lagrangean, discuss its symmetries and the corresponding conserved Noether 
charges. We reformulate the theory in hamiltonian formulation, and rederive the
(supersymmetry and internal) transformations generated by these charges. 
Super-Poincar\'{e} algebra is also realized in this formulation.
}}

\vfill 
\nit
\footnoterule 
\nit
{\footnotesize{$^\dag$ e-mail: tinosn@nikhef.nl \\
 }}
\np

\pagestyle{plain} 
\pagenumbering{arabic}

\section{Introduction \label{s1}} 

In recent years, there has been an increased interest in studying 
hydrodynamical systems from various points of view. The supersymmetric 
generalization of planar Chaplygin gas arising from supermembrane theory in 
(3 + 1)-dimensional space-time was proposed in \cite{jp,h}. In (1 + 1) 
dimensions, there is a large class of models. As an 
example we mention a supersymmetric fluid model which has been studied by 
Bergner and Jackiw \cite{jb}. Related 
developments in supersymmetrization of gas models can be found in \cite{m,az}.
 
We have recently proposed a supersymmetric theory of hydrodynamics in 
four-dimensional Minkowski space, for irrotational flow, that is, when 
the vorticity --the circulation in the motion of fluid around a fixed 
point-- is zero \cite{tjs}. However, general fluid motions are not 
irrotational. It is therefore, desirable to explore the possibility of 
extending the model to a wider class of flows, such as flow with 
non-vanishing vorticity. However, in the presence of vorticity, 
the current Chern-Simons term could provide an obstruction to constructing 
a lagrangean for fluid motion. This obstruction is absent if the 
hydrodynamical current is decomposed into three scalars in so-called 
Clebsch\footnote{Parametrization of any vector field $A_\gm$ in terms of 
three scalar potential $(\ga,\gb,\gg)$ is called Clebsch 
parametrization: $A_\gm = \der_\gm\ga + \gb\der_\gm\gg$} 
parametrization \cite{jackiw1}.

In a recent work \cite{tjs1} an alternative to the Clebsch decomposition of 
hydrodynamical currents in terms of complex potentials taking values in a 
K\"{a}hler manifold has been proposed. It introduces a general 
lagrangean density reproducing equations of motion of a perfect 
(dissipationless) relativistic fluid:
\begin{eqnarray}
\der^{\mu} T_{\mu\nu} &=& 0,\quad T_{\mu\nu} = p g_{\mu\nu} + 
(\varepsilon + p) u_{\mu} u_{\nu},\quad\der_\gm j^\gm = 0,\quad j^\gm = 
\gr u^\gm.
\label{continuity}
\end{eqnarray}
where $p$ is the pressure, $\varepsilon$ is the energy-density and 
$u^{\mu}$ is the velocity four-vector. The vanishing divergence of the fluid 
density current $j^{\mu}$ is the equation of continuity in hydrodynamics
which expresses the conservation of the fluid density $\gr$ during the flow, 
whilst the divergence-free energy-momentum tensor $T_{\mu\nu}$ expresses the 
conservation of energy and momentum. In addition to \eqref{continuity}, 
the \Kh\ parametrization allows a straightforward generalization to the 
case of superfields. Thus it can be used to construct a supersymmetric 
extension of relativistic fluid mechanics.

In this paper, we work out some properties of the
supersymmetric extension of fluid mechanics proposed in \cite{tjs1} and 
discuss in details various aspects of this supersymmetric field theory. 
In section \ref{s2}, we present the supersymmetric lagrangean in terms of 
superfields and work out its component form. In section \ref{s3} we 
discuss the internal symmetries of these lagrangeans in terms of 
Killing vectors, which represent infinitesimal symmetry transformations. 
Then we construct infinitesimal supersymmetry transformations of the fields 
appearing in the lagrangean. Using Noether's procedure, we construct the 
conserved quantities associated to these symmetries, such as supercharges, 
which are the generators of supersymmetry transformations, as well as the 
energy-momentum tensor from which the four-momentum is constructed. 
A canonical formulation of the theory in terms of a hamiltonian with a 
corresponding bracket structure is given in section \ref{s4}. 
Section \ref{s5} contains the conclusions.

\section{Supersymmetric lagrangeans \label{s2}} 

In this section, we construct a supersymmetric lagrangean, using the 
tensor calculus as described in \cite{AvanP}. Our aim is to 
arrive at a recipe which will allow us to write down a general 
supersymmetric theory, so that later we can apply the results to the 
special case of a supersymmetric extension of relativistic fluid mechanics. As supersymmetric models in four dimensions require the target 
manifold of scalar fields to be a \Kh\ manifold, we 
summarize here the most important results concerning \Kh\ manifolds relevant 
for later discussions; for more detailed and complete discussion 
on \Kh\ manifolds see \cite{zumino,Nakahara,Wells}.

A \Kh\ manifold is a complex manifold, parametrized locally by $N$ 
complex coordinates $z^\ga$ and their complex conjugates $\bz^\ugb$ 
($\ga, \ugb = 1,\dots, N$) on which a real line element can be 
defined by
\equ{
\text{d}s^2 = g_{\ga\ugb}\,\text{d}\bz^\ugb\,\text{d}z^\ga,
}   
with a hermitian metric $g_{\ga\ugb}$. The hermitian metric $g_{\ga\ugb}$ is 
said to be \Kh ian if the corresponding \Kh\ 2-form $\go = 
- i g_{\ga\ugb}\,\text{d}\bz^\ugb\wedge\text{d}z^\ga$ is closed,
\begin{eqnarray}
\text{d}\go &=& -\frac{i}{2}\Bigl(g_{\ga\ugb,\gg} - g_{\gg\ugb,\ga}\Bigl)
\text{d}z^\gg\wedge\text{d}z^\ga\wedge\text{d}\bz^\ugb \nonumber\\
[2mm]
& &
-\frac{i}{2}\Bigl(g_{\ga\ugb,\ugg}
- g_{\ga\ugg,\ugb}\Bigl)\text{d}z^\ga
\wedge\text{d}\bz^\ugb\wedge\text{d}\bz^\ugg  = 0
\labl{closed}
\end{eqnarray}
The comma denotes a derivative with respect to $z^\ga$ and $\bz^\ugb$.
The requirement \eqref{closed} is equivalent to the equations 
\equ{
g_{\ga\ugb,\gg} = g_{\gg\ugb,\ga},\quad g_{\ga\ugb,\ugg} = g_{\ga\ugg,\ugb}.
} 
From now on, we consider \Kh\ manifolds of complex dimension one, which are 
relevant for the following discussion. Locally, the metric can be derived from a scalar potential 
$K(z, \bz)$, the \Kh\ potential, as a second mixed derivate with 
respect to $z$ and $\bz$
\equ{
g_{z\bz} = K_{,z\bz}.
\labl{metric}
}
The complex connections $\gG^z_{zz}$ and curvature tensor $R_{z\bz z\bz}$ of a \Kh\ manifold are given by 
\equ{
\gG^z_{zz} = g^{z\bz}g_{\bz z,z},\quad R_{z\bz z\bz} = g_{z\bz,z\bz} - g_{\bz z,\bz}g^{z\bz}
g_{\bz z,\bz}.
\labl{conection}
}
where $g^{z\bz}$ is the inverse of the metric $g_{z\bz}$. 

Having briefly discussed the \Kh\ geometry, we now turn to the 
construction of the action. It is defined in terms of two sets of 
chiral\footnote{Our conventions for chiral spinors are given in the appendix}
superfields $\gF= (z, \eta_+, H)$, $\gL =  (s, \gc_+. h)$ and a real vector 
multiplet 
$V = ( C, \gps_\pm,\cZ, V_\gm,\gl_\pm, D)$. In terms of these multiplets, we 
propose the following action \cite{tjs1}
\equ{
S\, =\, \int d^4 x\,\cL,\quad\text{with}\quad\cL = \Bigl[ 
V\Bigl(K(\Phi,\bar{\Phi}) + \Lambda + \bar{\Lambda}\Bigl)  - \cF(V) \Bigl]_D 
\labl{action}
}
Here $\cF(V)$ is an analytic function of the real vector multiplet $V$. 
The component form of the action \eqref{action} after eliminating the 
auxiliary fields $D, \cH, h$, $H, \gc_+, \lambda_+$ and their complex 
conjugates reads
\begin{eqnarray}
\cL &=& V^\gm\Bigl( 2 \der_\gm N - i K_{,z}\der_\gm z + 
i K_{,\bz}\der_\gm\bz 
+ 2 i g_{z\bz}\,\bar{\eta}_-\gg_\gm\eta_+ +
\frac{i}{2} \cF^{\prime\prime\prime}(C)\bgps_+\gg_\gm\gps_-\Bigl)
\nonumber\\
[2mm]
& &
-\, 2 C \Bigl( g_{z\bz}\,\der_\gm z\partial^\gm\bz +  
g_{z\bz}\,\bar{\eta}_+\stackrel{\leftrightarrow}{\sDer}\eta_- 
- R_{z\bz z\bz}
\,\bar{\eta}_+\,\eta_+\,\bar{\eta}_-\,\eta_-\Bigl)  
\nonumber\\
[2mm]
& &
\labl{susyLag}
- \,\frac{1}{2} \cF^{\prime\prime}(C)\Bigl[ \partial_\gm 
C\partial^\gm C - 
V_{\gm}V^\gm + \bgps_+ \stackrel{\leftrightarrow}
{\sder}\gps_-\Bigl]  - \frac{2}{C}
g_{z\bz}\,\bgps_+\,\eta_+
\,\bgps_-\,\eta_-
\\
[2mm]
& &
 +  \,2 i g_{z\bz}\Bigl(\bgps_+\der\slashed z\eta_- -  
\bgps_-\der\slashed\bz\eta_+\Bigl) - \frac{1}{8} 
\cF^{\prime\prime\prime\prime}(C)
\bgps_+\,\gps_+\,
\bgps_-\,\gps_-
\nonumber.
\end{eqnarray}
In this expression we used the notation of the  
metric $g_{z\bz}$, connection $\gG^z_{zz}$ and curvature  
$R_{z\bz z\bz}$, given in \eqref{metric} and 
\eqref{conection} respectively. The primes are the derivatives of $\cF(C)$ 
w.r.t.\ its argument and $N = \text{Im}\,s$ is 
the imaginary\footnote{We take this 
opportunity to 
point out that $N$ is the $\text{Im}\,s$ and not $\text{Re}\,s$ as it 
appeared in \cite{tjs}.} component of $s$. The 
\Kh\ covariant derivative of a chiral 
spinor and the left-right arrow above the covariant derivative are are 
defined by
\begin{eqnarray}
\sDer\eta_- &=& \sder\eta_- + \bgG_{\bz\bz}^\bz\sder\bz
\eta_-,\quad\bgps_\pm\stackrel{\leftrightarrow}{\sder}\gps_\pm = 
\bgps_\pm\gg^\gm\der_\gm\gps_\pm - \der_\gm\bgps_\pm\gg^\gm\gps_\pm
\nonumber\\
[2mm]
\sDer\eta_+ &=& \sder\eta_+ +\gG_{zz}^z\sder z
\eta_+.
\end{eqnarray}

\section{Symmetries and currents\label{s3}} 

In the this section, we discuss the symmetries of the theory described by the 
action \eqref{action}, and the resulting conserved quantities. 
We first discuss internal symmetry, then we give the infinitesimal 
supersymmetry transformations. After that we construct the energy-momentum 
tensor following from the invariance of the action under translation.

Internal symmetries are realized as a set of holomorphic 
Killing vectors $R^i(z)$ and $\bR^i(\bz)$, which are a 
solutions of the Killing equation
\equ{
R^i_{\bz,z} + \bR^i_{z,\bz} = 0.
\labl{isom2}
} 
Here $R^i_\bz = g_{z\bz}R^{iz}$. The Killing vectors generate the isometries 
of the manifold corresponding to the coordinate transformations
\equ{
\gd z =\, z^\prime\, -\, z\, =\,\gTh^i\, \gd_i z\, = \gTh^iR^z_i(z),
\label{isometries}
} 
with $\gTh^i$ the parameters of the infinitesimal transformations. The 
isometries define a Lie algebra of the isometry group $\cG$, with structure constants $f_{ij}^{\;\;\;k}$ via the
Lie derivative by:
\begin{equation}
\lh {\cal L}_{R_i}[R_j] \rh^z\, =\,
R_i^{\, z} R^{\,z}_{j, z}\, -\,
R_j^{\, z} R^{\,z}_{i, z}\, =\, f_{ij}^{\;\;\;k}\,
R_k^{\, z}.
\labl{LiDer}
\end{equation} 
The infinitesimal transformations of other superfield components are 
found by requiring the isometries to commute with supersymmetry \cite{jv}
\equ{
\gd\eta_+ = \gTh^iR^z_{i,\,z}(z)\eta_+,\quad
\gd H = \gTh^i\Bigl(R^z_{i,\,z}(z)H - R^z_{i,\,zz}(z)\bar{\eta}_+\eta_+\Bigl).
\labl{isometries2}
}
If $\cG$ has a subgroup $\cH$ under which the transformations 
\eqref{isometries} are linear in $z$, we may visualize the manifold 
as a coset space $\cG/\cH$. A particularly simple example is provided by the coset 
space $SU(2)/U(1)$ with the \Kh\ potential 
\equ{
K(\bz,z) = \ln (1 + \bz z).
\label{2}
}
The explicit form of the Killing 
vectors \eqref{isometries} in the $SU(2)/U(1)$  case are 
\equ{
\gd z = \ge + i \gth z + \bar{\ge} z^2 = R^z(z), 
\labl{su2}
}
with $\gTh = (\gth,\ge, \bge)$. Here $\gth$ is the parameter of 
$U(1) \subset SU(2)$ phase transformation, and $(\bge, {\ge})$ are the complex 
parameters of the broken off-diagonal $SU(2)$ transformations. Under the 
isometry transformations \eqref{su2} the \Kh\ potential 
$K(z,\bz)$ is invariant up to the real part of a holomorphic function $F(z)$ 
transforming in the adjoint representation of the algebra \eqref{LiDer}:
\equ{
\gd K(\bz,z) = F(z) + \bF(\bz),\quad
 F(z;\gth,\bge) = \frac{i}{2}\, \gth + \bge z.
} 
The transformations of chiral superfield $\Lambda$ and the real vector superfield $V$ are obtained by 
requiring that the full action to be invariant under the isometry 
transformations \eqref{su2}:
\equ{
\gd\Lambda = - F(\gF),\quad\gd V = 0.
}
Observe, that the transformations of the scalar $N$ can therefore be 
written as 
\equ{ 
\gd N  = \frac{i}{2}\Bigl(F(z) - \bF(\bz)\Bigl).
\labl{isoN}
} 
The isometries \eqref{su2} can be obtained locally 
as gradients of a set of real Killing potentials, defined by 
\equ{
G(\gth,\ge,\bge) = i\Bigl(K_{,z}\,R(z) - F(z)\Bigl) = \frac{1}{2}\frac{\gth ( 1 - \bz z) + 
 2 i (\ge \bz - \bge z)}{1 + \bz z}. 
\label{11}
}
Indeed, the variations \eqref{isometries} are given by
\equ{
\gd z =  -i g^{z\bz}G_{,\bz}(\gth,\ge,\bge).
}

Let us now return to the case of general $K(z,\bz)$ and construct the 
charges. A set of conserved currents can be derived using the Noether 
procedure from the isometry transformations \eqref{isometries}, 
\eqref{isometries2} 
and \eqref{isoN} . The resulting currents are:
\begin{eqnarray}
J_\gm (G) &=& - 2 V_\gm\,G - 2 G_{;z}\Bigl(iC\der_\gm z - \bgps_-\gg_\gm\eta_+
\Bigl) 
+ 2 G_{;\bz}\Bigl(i C\der_\gm\bz + \bgps_+\gg_\gm\eta_-
\Bigl)\nonumber\\
[2mm]
& &
+ 4 iC G_{;\bz;z}\,\bar{\eta}_-\gg_\gm\eta_+,
\labl{isocurrent}
\end{eqnarray}
where the semicolon denotes a covariant derivative using the connection 
\eqref{conection}. These currents are divergence free 
\equ{
\der\cdot J = 0
\labl{divfree}
}
as it can be verified explicitly using the equations of motion. The field
equations derived from lagrangean \eqref{susyLag} read
\begin{eqnarray}
\cF^{\prime\prime}(C)\Box C &= & 2 g_{z\bz}
\Bigl(\der z\cdot\der\bz + \bar{\eta}_+\stackrel{\leftrightarrow}
{\sDer}\eta_- 
\Bigl) - \frac{1}{2} \cF^{\prime\prime\prime}(C)
\Bigl[ V^2 + 
(\der C)^2 \nonumber\\
& &- \bgps_+ \stackrel{\leftrightarrow}{\sder}
\gps_-\Bigl] - \frac{i}{2} \cF^{\prime\prime\prime\prime}(C)
\bgps_+\,V\Slashed\,\gps_-
\labl{C} + \frac{1}{8} \cF^{\prime\prime\prime\prime\prime}(C)
\bgps_+\,\gps_+\,
\bgps_-\,\gps_-\nonumber\\
& &- \frac{2}{C^2}\,g_{z\bz}\bgps_+\,
\eta_+\bgps_-\,\eta_- + 2 R_{z\bz z\bz}\bar{\eta}_+\,\eta_+\bar{\eta}_-\,
\eta_-
\\
[2mm] 
\cF^{\prime\prime}(C)V_\gm  + 2\der_\gm N&=& i \Bigl(K_{,z}\partial_\gm z - K_{,\bz}
\partial_\gm\bz  
- 2 g_{z\bz}\bar{\eta}_-\gg_\gm\eta_+ \Bigl) - \frac{i}{2}\cF^{\prime\prime\prime}(C)\bgps_+\gg_\gm\gps_- 
\labl{N}
\\
[2mm] 
\der\cdot V &=& 0,\labl{V}\\
[2mm]
-2 C g_{z\bz}\Box z &=&  2 ig_{z\bz}\,\bgps_-\sder\eta_+ + 2 ig_{z\bz,z}\,
\bgps_-\sder z\eta_+-
 2i g_{z\bz}\Bigl( V\cdot\der z  
+ \bar{\eta}_+\sder\gps_-\Bigl)\nonumber\\
[2mm]
& & 
+ g_{z\bz,\bz}\Bigl(2i\bar{\eta}_-V\Slashed\eta_+ - \frac{2}{C}\bgps_+\,\eta_+
\,\bgps_-\,\eta_- + 2i \bgps_+\sder z\eta_- \nonumber\\
[2mm]
& & 
- 4 C\bar{\eta}_-\sder\eta_+ + 2\bar{\eta}_+\sder C\eta_-\Bigl) + 
2 C g_{z\bz}\,\der C\cdot\der z + 2 Cg_{z\bz,z}\,\der z\cdot\der z
\nonumber\\
[2mm]
& & 
+ 2 C R_{z\bz z\bz,\bz}\bar{\eta}_+\eta_+\bar{\eta}_-\eta_- + 
 4 C g_{\bz z,\bz z}\bar{\eta}_+\sder z\eta_-
\labl{eqz}
\end{eqnarray}
for the bosonic fields, and
\begin{eqnarray}
\cF^{\prime\prime}(C)\sder\gps_+ &=& - \frac{1}{2} 
\cF^{\prime\prime\prime}(C)\Bigl(\partial\slashed C + i V\Slashed\Bigl)
\gps_+
- \frac{1}{4} \cF^{\prime\prime\prime\prime}(C)\gps_-\,\bgps_+\,
\gps_+
\nonumber\\
[2mm]
& &
- 2 g_{z\bz}\Bigl(\frac{1}{C}\eta_-\bgps_+ 
+ i \der\slashed\bz\Bigl)\eta_+,
\labl{ggz}
\\
[2mm]
4 C\sDer\eta_+ &=& - 2 \Bigl(\partial\slashed C 
- i V\Slashed + \frac{1}{C}\gps_-\bgps_+ \Bigl)\eta_+ 
- 2 i\sder z\gps_+ 
\nonumber\\
[2mm]
& &
\, + 
4 C g^{z\bz} R_{z\bz z\bz}\eta_-\bar{\eta}_+\eta_+,
\labl{gps}
\end{eqnarray}
for the fermionic ones. All other equations of motion for 
$(\bz,\gps_-,\eta_-)$ are obtained by complex 
conjugation of \eqref{eqz}, \eqref{ggz} and \eqref{gps} respectively. 
Expression \eqref{eqz} simplify considerably if we use the field equation 
\eqref{gps} for $\eta_+$
\begin{eqnarray}
-2 \cD\cdot(C\der z) &=&  2 i\bgps_-D\Slashed\eta_+ -
 2i \Bigl( V\cdot\der z  
+ \bar{\eta}_+\sder\gps_-\Bigl)\nonumber\\
[2mm]
& & 
+ 2 C g^{z\bz}R_{z\bz z\bz;\bz}\bar{\eta}_+\eta_+\bar{\eta}_-\eta_- + 
 4 C g^{\bz z}R_{\bz z\bz z}\bar{\eta}_+\sder z\eta_-.
\end{eqnarray}
Here the curly $\cD$ represents a \Kh\--covariant derivative:
\equ{
\cD_\gm (\der^\gm z) = \Box z + \der z\gG_{zz}^z\cdot\der z.
}
The conserved charges $q(G)$ are obtained from the divergence free currents 
\eqref{divfree}
\equ{
q(G) = \int d^3 x J^0(G).
\labl{charges}
} 

We now turn to the construction of supercharges. The infinitesimal supersymmetry transformations leaving the action 
\eqref{susyLag} invariant, with anti-commuting chiral 
spinor parameters $\ge_+$ and $\ge_-$ are
\equ{
\barr{ll} 
\gd C = \frac{i}{2}\bar{\ge}_+\gps_+ -  \frac{i}{2}
\bar{\ge}_-\gps_-, & \gd z = \bar{\ge}_+\eta_+ ,\\
 & \\
\gd\gps_+ = - \frac{1}{2}( V\Slashed  + i \der\slashed C )\ge_- & 
\gd\gps_- =  - \frac{1}{2}(V\Slashed - i \der\slashed C )
\ge_+, \\ 
 & \\  
\gd V_{\mu} = \bar{\ge}_+\gs_{\gm\gn}\partial^\gn\gps_+ + 
\bar{\ge}_-\,\gs_{\gm\gn}\partial^\gn\gps_- &\gd\bz = \bar{\ge}_-\eta_- ,\\
& \\
\gd N = \frac{1}{4}\cF^{\prime\prime}(C)(\bar{\ge}_+\gps_+ + 
\bar{\ge}_-\gps_-) +
\frac{i}{2}(\bar{\ge}_+ K_{,z}\eta_+ - \bar{\ge}_- K_{,\bz}\eta_-), \\
 & \\
\gd\eta_+ = \frac{1}{2}\Bigl(\der\slashed z\ge_- + \ge_+\frac{1}{2 C} 
g^{z\bz}\,(i \bgps_+\eta_+ + 2 C g_{z\bz,z}\bar{\eta}_+\eta_+)\Bigl)
, \\
 & \\
\gd\eta_- = \frac{1}{2}\Bigl(\der\slashed\bz\ge_+ + \ge_-\frac{1}{2 C} 
g^{z\bz}\,(- i \bgps_-\eta_- + 2 C g_{z\bz,\bz}\bar{\eta}_-\eta_-)\Bigl). 
\earr
\label{susyva}
}
Under these transformations the variation of the lagrangean is a total 
derivative:
\equ{
\gd\cL = \der_\gm \Bigl(\frac{i}{2}\bar{\ge}_+B^\gm_+ - \frac{i}{2}\bar{\ge}_-B^\gm_-\Bigl),
}
with the vector-spinor $B^\gm_{\,\pm}$ given, modulo equations of motion, by
\begin{eqnarray}
B^\gm_{\,+} &\simeq& 2 g_{z\bz}\gg^\gm\,\eta_-\bgps_+\,\eta_+\, 
+ 2 i C g_{z\bz}\,\der\slashed\bz\gg^\gm\,\eta_+ - 
\frac{1}{2}\cF^{\prime\prime}(C)\gg^\gm\Bigl(\der\slashed C 
+ 
i V\Slashed\Bigl)\gps_+
\nonumber\\
[2mm]
& & 
-\, \frac{1}{2}\cF^{\prime\prime\prime}(C)\gg^\gm\,\gps_-\,\bgps_+\gps_+
\labl{vspinor}
\\
[2mm]
B^\gm_{\,-} &\simeq& 2 g_{z\bz}\gg^\gm\,\eta_+\,\bgps_-\,
\eta_-
- i 2 C g_{z\bz}\,\der\slashed z\gg^\gm\eta_- - 
\frac{1}{2}\cF^{\prime\prime}(C)\gg^\gm\Bigl(\der\slashed C - 
i V\Slashed\Bigl)\gps_-
\nonumber\\
[2mm]
& & 
-\, \frac{1}{2}\cF^{\prime\prime\prime}(C)\gg^\gm\gps_+\,\bgps_-\gps_-\nonumber.
\end{eqnarray}
where the similarity sign $\simeq$ in \eqref{vspinor} 
signifies that the vector-spinors are given up to equations of 
motion. The supercurrents $S^\gm_{\,\pm}$ following from the 
invariance of the action \eqref{action} under supersymmetry variations 
\eqref{susyva} are obtained using standard methods. By the usual Noether 
theorem, one finds for the supercurrent (and its hermitian conjugate) 
in terms of the variations of the complete set of fields $A$:
\equ{
\frac{i}{2}\bar{\ge}_+S^\gm_{\,+} - \frac{i}{2}\bar{\ge}_-S^\gm_{\,-} = 
\frac{i}{2}\bar{\ge}_+B^\gm_{\,+} - \frac{i}{2}\bar{\ge}_-B^\gm_{\,-} 
- \frac{\gd\cL}{\gd\der_\gm A}
\gd A,
}
with $B^\gm_{\,\pm}$ given in \eqref{vspinor}. A little work 
reveals that
\begin{eqnarray}
S_{\gm +} &= & 4 C g_{z\bz}\gg_\gm\,\eta_-\bgps_+\,\eta_+
- 4i C g_{z\bz}
\der\slashed\bz\gg_\gm\eta_+ + 
\cF^{\prime\prime}(C)(\der\slashed C + i V\Slashed )\gg_\gm\gps_+
\nonumber\\
[2mm]
& &
- \frac{1}{2}\cF^{\prime\prime\prime}(C)\gg_\gm\gps_-\,\bgps_+\,\gps_+
- 2 i C g_{z\bz,z}\gg_\gm\eta_-\,\bar{\eta}_+\eta_+,
\labl{supercurr}
\nonumber\\
[2mm]
S_{\gm -} &= &
4 g_{z\bz}\gg_\gm\eta_+\,\bgps_-\,\eta_- + 4i C g_{z\bz}
\der\slashed z\gg_\gm\eta_- + 
\cF^{\prime\prime}(C)(\der\slashed C - i V\Slashed )
\gg_\gm\gps_-
\nonumber\\
[2mm]
& &
- \frac{1}{2}\cF^{\prime\prime\prime}(C)\gg_\gm\gps_+
\bgps_-\gps_-
+ 2 i C 
g_{\bz z,\bz}\gg_\gm\eta_+\,\bar{\eta}_-\eta_-.
\end{eqnarray}  
The supercurrents and its hermitian conjugate are separately conserved:
\equ{
\der_\gm S^\gm_\pm = 0,
\labl{scurrents}
} 
as can be verified by use of equations of motion. From these currents one 
constructs the conserved spinors supercharges
\equ{
Q_\pm = \int d^3 x S^0_\pm,
\labl{scharges}
}
which are the generators of supersymmetry transformations. 

Since supersymmetric field theories are translationally invariant, 
the theory described by lagrangean \eqref{susyLag} conserves energy-momentum. 
By the Noether procedure, translation invariance leads to non-symmetric 
currents  $\gTh_{\gm\gn}$ defined by
\equ{
\gTh_{\gm\gn} = - \frac{\gd\cL}{\gd(\der_\gm A_i)}\der^\gn A_i + 
g_{\gm\gn}\cL,
\labl{gTh}
}
where $\cL$ is the lagrangean \eqref{susyLag}, and the sum is taken over 
the various fields of the theory. The symmetrized version of the energy-momentum tensor $T_{\gm\gn}$ 
is obtained by the addition of an improvement term $\gO_{\gm\gn}$ which is the 
anti-symmetric part of \eqref{gTh}
\begin{eqnarray}
T_{\gm\gn} &=& \cF^{\prime\prime}(C)\Bigl(\partial_\gm C\partial_\gn C +
V_{\gm}V_\gn + \frac{1}{4}\bgps_+\gg_{\{\gm} \stackrel{\leftrightarrow}
{\der}_{\gn\}}\gps_-\Bigl) + \frac{i}{4}\cF^{\prime\prime\prime}(C)
\bgps_+V_{\{\gm}\gg_{\gn\}}\gps_- \nonumber\\
[2mm]
& &
+ 2 C g_{z\bz}\Bigl(\partial_\gm z\partial_\gn\bz + \partial_\gm\bz\partial_\gn z +
\frac{1}{2}\bar{\eta}_+\gg_{\{\gm} \stackrel{\leftrightarrow}
{\der}_{\gn\}}\eta_-\Bigl) 
+  i g_{z\bz}\bar{\eta}_-V_{\{\gm}\gg_{\gn\}}\eta_+\nonumber\\
[2mm]
& &
- \, \Bigl[ i g_{z\bz}\bgps_+\der_{\{\gn} z\gg_{\gm\}}\eta_- + 
\text{h.c.}\Bigl] -
\Bigl[C g_{z\bz,z}\bar{\eta}_+\der_{\{\gn} z\gg_{\gm\}}\eta_- +
 \text{h.c.}\Bigl] + g_{\gm\gn}\cL,
\labl{energym}
\end{eqnarray}
Here $\{\gm, \gn\}$ denotes the symmetrized expression.
The conservation of the energy-momentum tensor follows up on using the 
field equations. From this energy-momentum tensor one construct the 
conserved four-momentum
\equ{
P_{\mu} = \int d^3 x\,T_{\mu 0}.
\labl{4momentum}
}
In the following section, we construct the explicit expressions for the 
supercharges \eqref{scharges} and four-momentum vector \eqref{4momentum}.

\section{Canonical analysis\label{s4}} 

In this section, we show that the supercharges $Q_\pm$ satisfy the 
supersymmetry algebra and that they generate 
the supersymmetry transformations \eqref{susyva} as well as the space-time 
translations on the fields. As this action of the supersymmetry 
algebra in terms of $Q_\pm$ requires the use of canonical 
variables and hamiltonian equations of motion, we first present 
a canonical formulation of the theory and describe the dynamics in 
terms of phase-space coordinates and the hamiltonian.  However in this 
formalism, the fermionic momenta turn out not to be independent degrees of 
freedom, as they are constrained to fermionic fields themselves. To 
eliminate these constraints we introduce Poisson-Dirac brackets, defined 
as the Poisson brackets from which the second class 
constraints have been projected out.

We now present details of this analysis. The canonical momenta of the 
theory are defined by
\begin{eqnarray}
\pi_C &=& \frac{\gd \cL}{\gd \dot{C}}\, = \cF^{\,\prime\prime}(C)
 \dot{C},\quad\pi_N = \frac{\gd \cL}{\gd \dot{N}} = 2 V^0,\nonumber\\
[2mm]
\pi_z &=& \frac{\gd \cL}{\gd \dot{z}} = i K_{,z} V_0 + 2 C 
g_{z\bz}\dot{\bz} +
2 i g_{z\bz}\bgps_+\gg^0\eta_- + 2 C g_{z\bz,z}\bar{\eta}_+\gg^0\eta_-,\nonumber\\
[2mm]
\bar{\pi}_\bz &=& \frac{\gd \cL}{\gd \dot{\bz}} = - i K_\bz V_0 + 2 C 
g_{z\bz}\dot{z} 
- 2 i g_{z\bz}\bgps_-\gg^0\eta_+ + 2 C g_{z\bz,\bz}\bar{\eta}_-\gg^0\eta_+\nonumber\\
[2mm]
\pi_{\gps_\pm} &=& \gg_0\, \frac{\gd \cL}{\gd \dot{\bgps}_\mp} = 
 \frac{1}{2}\cF^{\,\prime\prime}(C)\gps_\pm,\quad 
\pi_{\eta_\pm} = \gg_0\frac{\gd \cL}{\gd \dot{\bar{\eta}}_\mp} =
 2 C g_{z\bz}\eta_\pm.
\label{12}
\end{eqnarray}
Here we included $\gg_0$ in the definition of the fermionic 
momenta so that the momenta of Majorana variables are Majorana themselves 
as well. Clearly, the last 
two equations of \eqref{12} are second-class constraints, expressing the 
fermionic momenta $(\pi_{\gps_\pm},  \pi_{\eta_\pm})$ 
in terms of fermionic fields:
\begin{eqnarray}
\gc_{\gps_\pm} &=& \pi_{\gps_\pm} - \frac{1}{2}
\cF^{\prime\prime}(C)\gps_\pm \simeq 0,\quad \gc_{\eta_\pm} = 
\pi_{\eta_\pm} - 
2 C g_{z\bz}\eta_\pm \simeq 0
\label{constraints}
\end{eqnarray}
The similarity sign $\simeq$ in last equality of \eqref{constraints} 
signifies that the constraints are defined only on a subset 
(the physical shell) of the full phase space. In this extended phase space, 
the equal-time Poisson brackets of the theory are defined by
\begin{eqnarray}
\left\{ \pi_{\eta_\pm}({\bf r}), \bar{\eta}_\mp({\bf r}^{\prime}) \right\} &=& 
\left\{\pi_{\gps_\pm}({\bf r}),\bgz_\mp({\bf r}^{\prime}\right\} = 
\left\{\gps_\pm({\bf r}^{\prime}),\bar{\pi}_{\gz_\mp}({\bf r})\right\} = 
\gg^0P_\mp\,\gd^3({\bf r} - {\bf r}^{\prime})\nonumber\\
[3mm]
\left\{ \eta_\mp({\bf r}^{\prime}),\bar{\pi}_{\eta_\pm}({\bf r}) 
\right\} &=& 
\gg^0P_\pm\,\gd^3({\bf r} - {\bf r}^{\prime}),\quad \left\{ N({\bf r}), 
 \pi_N({\bf r}^{\prime}) \right\} =\gd^3({\bf r} - {\bf r}^{\prime}),\nonumber\\
[3mm]
\left\{ z({\bf r}), \pi_z({\bf r}^{\prime}) \right\} &=& 
\left\{ \bz({\bf r}), \bar{\pi}_\bz({\bf r}^{\prime}) \right\} = 
\left\{ C({\bf r}), \pi_C({\bf r}^{\prime}) \right\} =
 \gd^3({\bf r} - {\bf r}^{\prime}),
\end{eqnarray}
where $P_\pm = \frac{1}{2}(1 \pm \gg_5)$ are the left- and right-handed chiral
projection operators. 

In order to describe the canonical dynamics on the reduced 
phase-space determined by the constraint equations \eqref{constraints}, 
we introduce Poisson-Dirac brackets
\equ{
\left\{ A, B \right\}^* = \left\{ A, B \right\} -
 \left\{ A, \chi_i \right\} C^{-1}_{ij} \left\{ \chi_j, B \right\},
\label{21}
}
where $C^{-1}_{ij}$ is the inverse of the matrix of constraint brackets
\begin{eqnarray} 
C_{ij} &=& \{\gc_i({\bf r},t),\bgc_j({\bf r}^{\prime},t)\}
\\
[2mm]
&=&
- \left(
\mtrx{
0&\cF^{\prime\prime}(C)\gg^0 P_-&0&0\\
\cF^{\prime\prime}(C)\gg^0P_+&0&0&0\\
0&0&0&4 C g_{z\bz}\gg^0P_-\\
0&0&4 C g_{z\bz}\gg^0P_+&0
}
\right)\gd^3({\bf r} - {\bf r}^{\prime})\nonumber.
\end{eqnarray}
Applying this prescription, we obtain the full set of non-zero Poisson-Dirac 
brackets of our theory:
\begin{eqnarray}
\left\{ z({\bf r}),\pi_z({\bf r}^\prime)\right\}^* &=& \left\{\bz({\bf r}),
\bar{\pi}_\bz({\bf r}^\prime)  \right\}^* = 
\left\{ C,({\bf r}) \pi_C({\bf r}^\prime) \right\}^* = 
\gd^3({\bf r} - {\bf r}^{\prime}),\nonumber\\
[2mm]
\left\{ \pi_C({\bf r}), \bgps_\pm ({\bf r}^\prime)\right\}^* &=& 
  \frac{\cF^{\prime\prime\prime}(C)}{2\cF^{\prime\prime} (C)}
\bgps_\pm\gd^3({\bf r} - {\bf r}^{\prime}),\quad\left\{ \pi_z({\bf r}), \bar{\eta}_\pm ({\bf r}^\prime)\right\}^* = 
\frac{1}
{2}\gG^z_{zz}\,\bar{\eta}_\pm\gd^3({\bf r} - {\bf r}^{\prime})
,\nonumber\\
[2mm] 
\left\{ \bar{\pi}_\bz({\bf r}), \bar{\eta}_\pm({\bf r}^\prime)\right\}^* &=& 
 \frac{1}
{2}\gG^\bz_{\bz\bz}\,\bar{\eta}_\pm\gd^3({\bf r} - {\bf r}^{\prime})
,\quad\left\{ \eta_\pm({\bf r}), \bar{\eta}_\mp({\bf r}^\prime)\right\}^* =
\frac{1}{4C}g^{z\bz}\gg^0P_\mp\gd^3({\bf r} - {\bf r}^{\prime})
,\nonumber\\
[2mm]
\left\{ \gps_\pm({\bf r}), \bgps_\mp({\bf r}^\prime)\right\}^*&=& 
\frac{1}{2\cF^{\prime\prime}(C)}\gg^0P_\mp\gd^3({\bf r} - {\bf r}^{\prime}),
\\
[2mm]
\left\{ \eta_\pm({\bf r}), \pi_z({\bf r}^\prime)\right\}^*&= &
- \frac{1}{2}\gG^z_{zz}\eta_\pm\gd^3({\bf r} - {\bf r}^{\prime}),\quad\left\{ \eta_\pm({\bf r}), \bar{\pi}_\bz({\bf r}^\prime)\right\}^* =
- \frac{1}{2}\gG^\bz_{\bz\bz}\eta_\pm\gd^3({\bf r} - {\bf r}^{\prime}),\nonumber\\
[2mm]
\left\{ N({\bf r}),  \pi_N({\bf r}^\prime) \right\}^* 
 &=& \gd^3({\bf r} - {\bf r}^{\prime}),\quad\left\{ \gps_\pm({\bf r}), \pi_C({\bf r}^\prime)\right\}^* =
- \frac{\cF^{\prime\prime\prime}(C)}{2\cF^{\prime\prime}(C)}\gps_\pm\gd^3({\bf r} - {\bf r}^{\prime})\nonumber
\labl{Diracprackets2}
\end{eqnarray} 
The Poisson-Dirac brackets of Noether 
charges \eqref{charges} generate the isometries transformations, as defined 
in \eqref{isometries}, \eqref{isometries2} and \eqref{isoN}
\equ{
\gd_M A = \left\{q, A\right\}^*,
} 
where the canonical Noether charges \eqref{charges} is 
\begin{eqnarray}
q[G] &=& \int d^3 x\,\Bigl[\pi_N G - g^{z\bz}G_z\Bigl(i \bar{\pi}_\bz + 
\frac{1}{2}\,K_\bz \pi_N\Bigl) + g^{z\bz}G_\bz\Bigl(i \pi_z - 
\frac{1}{2}\,K_z \pi_N\Bigl)\nonumber\\
[2mm]
& &
- 2i C \Bigl(G_z\,\bgG_{\bz\bz}^\bz + G_\bz\,\gG_{z z}^z - 2 
G_{z\bz}\Bigl)\bar{\eta}_-\gg_0\eta_+\Bigl].
\end{eqnarray}
One may check the closure of the algebra of conserved charges by computing the Poisson-Dirac brackets of such two charges. After a long calculation, one finds that the 
result has the structure of a Poisson bracket on the 2-d manifold spanned 
by $(\bz,z)$.:
\equ{
\left\{ q[G^{(1)}], q[G^{(2)}] \right\}^* =  q[G^{(3)}],\quad\text{with}\quad  G^{(3)} = i\, g^{z\bz} \lh G^{(1)}_{z} G^{(2)}_{\bz} - 
 G^{(1)}_{\bz} G^{(2)}_{z} \rh.
\label{2.34}
}

The canonical hamiltonian, obtained from lagrangean \eqref{susyLag} by 
Legendre transformation, reads
\begin{eqnarray}
H &=&\int d^3 x\Bigl[ \frac{1}{2\cF^{\prime\prime}(C)}\pi_C^2 + 
\Bigl(\frac{1}{8}
\cF^{\prime\prime}(C) + \frac{1}{8C}g^{z\bz}\,K_{,z}K_{,\bz}\Bigl)\pi_N^2  + 
\frac{1}{2C}g^{z\bz}\pi_z\bar{\pi}_\bz \nonumber\\
[2mm]
& & 
-\, i \pi_N\Bigl(\frac{1}{4}\cF^{\prime\prime\prime}(C)\bgps_+ \gg_0 \gps_- +
g_{z\bz}\bar{\eta}_-\gg_0\eta_+ \Bigl) + \frac{1}{8}\cF^{\prime\prime\prime\prime}(C)
 \bgps_+\, \gps_+\,\bgps_-\, \gps_- \nonumber\\
[2mm]
& &
+\,\frac{1}{2}\cF^{\prime\prime}(C) \Bigl((\vec{\nabla} C)^2 + \vec{V}^2 + 
 + \bgps_+ \stackrel{\leftrightarrow}{\nabla\Slashed} \gps_- \Bigl) + 
2 C g_{z\bz}\Bigl(\vec{\nabla}z\vec{\nabla}\bz + 
\bar{\eta}_+ \stackrel{\leftrightarrow}{\nabla\Slashed} \eta_-\Bigl)
\nonumber\\
[2mm]
& &
- 2 C\Bigl( R_{z\bz z\bz} - \gG^z_{zz}\,g_{z\bz,\bz}
\Bigl)\bar{\eta}_+\,\eta_+\,\bar{\eta}_-\,\eta_- 
+ \frac{2}{C}g_{z\bz}\bgps_+\,\gg^0\,\eta_-\,\bgps_-\,\gg^0\eta_+ + 
\nonumber\\
[2mm]
& & 
\,+\frac{2}{C}g_{z\bz}\bgps_+\,\eta_+\,\bgps_-\,\eta_-
+ \Bigl\{ -\frac{i}{C}\bgps_+\gg^0\eta_-\bar{\pi}_\bz + \text{h.c}\Bigl\}
+ i g_{z\bz,\bz}\bar{\eta}_-\,\eta_-\,\bgps_+\,\eta_+ \nonumber\\
[2mm]
& & 
\,+\Bigl\{ \frac{i K_{,z}}{4 C g_{z\bz}}\pi_N\bar{\pi}_\bz + \text{h.c}\Bigl\} -
\Bigl\{  \frac{K_{,z}}{2 C}\pi_N\bgps_-\gg^0\eta_+ + \text{h.c}\Bigl\}
- i g_{z\bz,z}\bar{\eta}_+\,\eta_+\,\bgps_-\,\eta_- \nonumber\\
[2mm]
& & 
\,- \Bigl\{2 i g_{z\bz}\bar{\eta}_-\vec{\nabla}\Slashed z\gps_+ + 
2 C g_{z\bz,z}\bar{\eta}_+\vec{\nabla}\Slashed z\eta_- + 
\gG^\bz_{\bz\bz}\pi_z\,\bar{\eta}_-\,\gg^0\eta_+ + 
\text{h.c}\Bigl\}\Bigl].
\labl{susyHam}
\end{eqnarray}
In this expression we have used for the 3-dimensional contraction 
$\nabla\Slashed = \vec{\gg} \cdot \vec{\nabla}$ a notation analogous to the 
4-dimensional one. After a long and tedious calculation one finds that 
brackets with the hamiltonian reproduce all the field equations we derived 
earlier from the lagrangean \eqref{susyLag}:
\equ{
\der_0 A = \left\{ A, H\right\}^*.
}

We now turn to the construction of the canonical super-Poincar\'{e} 
algebra. First we construct the canonical expressions for the energy-momentum 
vector \eqref{4momentum} and the supercharges $Q_\pm$ 
\eqref{scharges}. For the four-momentum vector we find the result
\begin{eqnarray}
P_0 &=& \int d^3 x\cH = H\nonumber\\
[2mm]
P_i &=& \int d^3 x\Bigl[-\pi_C\nabla_iC -\pi_N\nabla_iN - 
-\frac{1}{2}
\cF^{\prime\prime}(C)\bgps_+\gg_0
\stackrel{\leftrightarrow}{\nabla\Slashed} \gps_- - 2 C g_{z\bz}\bar{\eta}_+\gg_0\stackrel{\leftrightarrow}{\nabla\Slashed}
\eta_-\nonumber\\
[2mm]
& &
\, 
-
\Bigl\{\frac{1}{8C}K_{,\bz}\pi_N\,\bar{\eta}_- \gg_i\gps_+ + \pi_z\Bigl(\nabla_iz - \frac{i}{4C}\,\bar{\eta}_+\gg_i\gps_-\Bigl) + \nonumber\\
[2mm]
& &
+ \frac{i}{2}\,g_{z\bz,z}\,(\bar{\eta}_+\gg_i\gps_-)
\,(\bar{\eta}_+\gg_0\eta_-) + \text{h.c}\Bigl\}\Bigl].
\end{eqnarray}
It generates space time translations on the fields $A$: 
\equ{
\der_\gm A = \left\{ A, P_\gm\right\}^*.
\labl{26}
}
The phase-space supercharges $Q_\pm$ are obtained directly 
from the supercurrents \eqref{supercurr}, which reads explicitly
\begin{eqnarray}
Q_+ &=&\int d^3{\bf r}\Bigl[ \Bigl(\cF^{\prime\prime}(C)\nabla\Slashed 
C -2 i\nabla\Slashed N + K_{,\bz}\nabla\Slashed\bz -  
K_{,z}\nabla\Slashed z\Bigl)\gg_0\gps_+ \nonumber\\
[2mm]
& & 
+ \frac{1}{4}\cF^{\prime\prime\prime\prime}(C)\gg_0\gps_-\,\bgps_+\,\gps_+  
+ \Bigl(\pi_C - \frac{i}{2}\cF^{\prime\prime}(C)\pi_N\Bigl)\gps_+ +
\Bigl(K_{,z}\pi_N - 2 i \pi_z\Bigl)\eta_+\nonumber\\
[2mm]
& & 
- 4 i C g_{z\bz}\nabla\Slashed\bz\gg_0\eta_+ - g_{z\bz}\gg_0\eta_-\bgps_+\eta_+ \Bigl] 
\labl{spinorch}
\\
[2mm]
Q_- &=& 
\int d^3{\bf r}\Bigl[\Bigl(\cF^{\prime\prime}(C)\nabla\Slashed 
C + 2 i\nabla\Slashed N + K_{,z}\nabla\Slashed z -  
K_{,\bz}\nabla\Slashed\bz\Bigl)\gg_0\gps_- + \nonumber\\
[2mm]
& & 
+ \frac{1}{4}\cF^{\prime\prime\prime\prime}(C)\gg_0\gps_+\,\bgps_-\,\gps_-
 + \Bigl(\pi_C + \frac{i}{2}\cF^{\prime\prime}(C)\pi_N \Bigl)\gps_- + 
\Bigl(K_{,\bz}\pi_N + 2 i \bar{\pi}_\bz\Bigl)\eta_-\nonumber\\
[2mm]
& &
+ 4 i C g_{z\bz}\nabla\Slashed z\gg_0\eta_- - 
g_{z\bz}\gg_0\eta_+\bgps_-\eta_-\Bigl]\nonumber.
\end{eqnarray} 
It is now straightforwards to generate the infinitesimal supersymmetry 
transformation \eqref{susyva} by  
Poisson-Dirac bracketing with the spinor charges \eqref{spinorch}
\equ{
\gd(\ge_\pm ) A = \pm\frac{i}{2}\left\{ A,\bar{\ge}_\pm\,Q_\pm
\right\}^*.
\labl{piosscharg}
}
These results show that the spinor charges $Q_\pm$ give the correct 
supersymmetry transformation of all the fields in the theory. The commutator 
of two such transformations with parameters $\ge^1$, $\ge^2$ indeed gives 
a translation 
with parameter $\gx^\gm = \frac{1}{2}\bge\,^2_+\gg^\gm\ge^1_- + 
\text{h.c.}$:
\equ{
 [ \gd(\ge^1),\gd(\ge^2)]A = \gx^\gm\der_\gm A.
\labl{commut}
}
This can be verified explicitly by using the Poisson-Dirac 
brackets. Then equation \eqref{commut} becomes:
\equ{
 \pm\frac{i}{2}\left\{\left\{ A,\bge\,^2_{\pm}\,Q_{\pm}\right\}^*,
\bge\,^1_{\pm}\,Q_{\pm}\right\}^* - ( \ge^2\leftrightarrow
\ge^1 ) =  \gx^\gm\der_\gm A
\labl{commut1}
}
up to terms which vanish on-shell. 
By rearranging the terms in \eqref{commut1} using the Jacobi 
identity, we therefore have
\equ{
\pm\frac{i}{2}\left\{A, \left\{\bge\,^2_{\pm}\,Q_{\pm},
\bge\,^1_{\pm}\,Q_{\pm} 
\right\}^*
\right\}^* = 
\gx^\gm\left\{ A, P_\gm\right\}^*,
}
for any $A$. It follows that
\equ{
\pm\frac{i}{2}\left\{\bge\,^2_{\pm}\,Q_{\pm},\bge\,^1_{\pm}\,Q_{\pm}\right\}^* = \gx^\gm P_\gm
\labl{commut2}
}
Now, by expanding out equation \eqref{commut2}, one obtains the supersymmetry algebra relations
\begin{eqnarray}
\left\{ Q_\pm,\bar{Q}_\mp \right\}^* &=& 2 P\Slashed,\quad
\left\{ Q_\pm, P_\gm \right\}^* = 0.
\label{25}
\end{eqnarray}
The last equation of \eqref{25} follow immediately from \eqref{26}. The 
brackets structure shows Poincar\'{e} supersymmetry to be realized also in 
the canonical formulation of the theory.

\section{Summary and discussion \label{s5}}

In this paper we have presented a complete formulation of a particular 
supersymmetric field theory on an arbitrary \Kh\ manifold. We have started 
with the construction of the lagrangeans then, discussed various aspects 
of these general constructions. The supercurrents and the energy-momentum 
tensor following from the invariance of the action under supersymmetry and 
translation are constructed. Assuming that there exists an 
isometry group $G$ which leaves the lagrangean invariant, we have 
constructed the corresponding conserved isometry currents, in terms of 
Killing vector $R^z(z)$.
   
Next, we discussed in details the canonical formulation of the 
theory in terms of hamiltonian. We have explicitly constructed the canonical 
supercharges. We have shown that these supercharges generate the 
supersymmetry transformations, and satisfy the standard super-Poincar\'{e} 
algebra. 

To get the hydrodynamical interpretation of the models, one has to relate 
the fields in our model to the 
particle number density $\gr$ and the velocity four-vector $u_\gm$, which is 
time-like $u^2_\gm = -1$. In particular, in the limit in which all fermion 
fields vanish, we have to identify the vector component $V_\gm$ with 
the particle number density 
$\gr$ and  $u_\gm$ as in \eqref{continuity}. This implies that $V_\gm$ is 
time-like four vector. However, this is not sufficient for this field 
theory to describe  a relativistic model of hydrodynamics. We have to 
show that this identification is consistent with the field equation 
\eqref{V} which is the relativistic equation of continuity. Finally, to 
complete the hydrodynamical interpretation of the models describe by 
\eqref{susyLag}, it should be possible to write the bosonic part of the 
energy-momentum tensor 
\eqref{energym} in standard form \eqref{continuity}.  The two energy-momentum 
tensors \eqref{continuity} and \eqref{energym} can be equivalent only 
in the hydrodynamical regime $(C,\eta_\pm\rightarrow 0)$ of the model. This 
analysis has been discussed in \cite{tjs1}.

\section*{Acknowledgements}

I would like to thank J.W.\ van Holten and S. Groot Nibbelink for very 
useful and pleasant discussions during this work, which is part of 
the research programme Theoretical Subatomic Physics (FP52) 
of the Foundation for Fundamental Research of 
Matter (FOM) and the Netherlands Organization for Scientific Research (NWO).

\appendix

\section{Appendix: Notations and conventions}

In this appendix we collect the conventions we used in this paper. 
Symmetrization of objects enclosed is denoted by braces 
$\left\{\dots\right\}$, anti-symmetrization by the square brackets 
$[\dots ]$; the total weight of such (anti-)symmetrization is always unit.

In this paper, we reformulated the theory in a hamiltonian formulation. 
To avoid a confusion, we have used the braces $\left\{,\right\}$ to denote
the Poisson brackets, $\left\{,\right\}^*$ to denote Poisson-Dirac brackets.

The Minkowski metric $g_{\gm\gn}$ has a signature $(-1,+1,+1,+1)$. Our conventions for chiral spinors 
\equ{
\gS_\pm = \frac{1 \pm \gg_5}{2}\gS_\pm,\quad
\gS_\pm\equiv (\gl_\pm,\gps_\pm,\eta_\pm,\gc_\pm )
}
are such, that $\gg_5 \gS_\pm = \pm\gS_\pm$ and 
$\bar{\gS}_\pm \gg_5 = \pm\bar{\gS}_\pm$; charge conjugations acts as 
$\gS_\pm = C \bar{\gS}_\pm^T$, where $\bar{\gS}_\pm = 
i \gS_\mp^{\dagger} \gg_0$. It should also be noted, that the euclidean 
$\gg_ 4 = i \gg_0$ is hermitean, hence $\gg_0$ is anti-hermitean. The charge-conjugation operator $C$ defined in the spinor space satisfies  
the properties: 
\equ{
C\, =\, C^{\dagger}\, =\, C^{-1}\, =\, -\, C^T,\quad 
C \gg_{\gm} C^{-1}\, =\, -\, \gg_{\gm}^T
}
where the superscript $T$ denotes transposition in spinor space.  From this it follows, that 
\equ{
C \gg_5 C^{-1}\, =\, \gg_5^T,\quad C \gs_{\gm\gn} C^{-1}\, =\, 
 - \gs_{\gm\gn}^T\quad\text{with}\quad\gs_{\gm\gn}\, =\, \frac{1}{4}\, \left[ \gg_{\gm}, \gg_{\gn} \right]. 
}

With these definitions one can show by taking the transposition of a scalar 
the following identities:
\begin{eqnarray}
\bar{\gS}_\pm\Id\gD_\pm &=& \bgD_\pm\,\Id\,\gS_\pm,\quad
~~~~\bar{\gS}_\pm\gg_\gm\,\gD_\pm = - \bgD_\pm\,\gg_\gm\,\gS_\pm,
\nonumber\\
[2mm]
\bar{\gS}_\pm\gg_5\gD_\pm &=& \bgD_\pm\,\gg_5\,\gS_\pm,\quad
\bar{\gS}_\pm\,\gs_{\gm\gn}\,\gD_\pm = - \bgD_\pm\,
\gs_{\gm\gn}\,\gS_\pm.
\end{eqnarray}
Hermitian conjugation on bispinor reverses by definitions the order of the 
spinors $(\gps\,\gc )^{\,\dag} = \gc^{\,\dag}\,\gps^{\,\dag}$ with no minus 
sign. Using this one can show that $\text{h.c.}$ replaces $i\leftrightarrow -i$ and 
$+\leftrightarrow -$:
\begin{eqnarray}
(\bgS_{\pm}\,\gD_{\pm})^{\,\dag} =
\bar{\gS}_\mp\,\gD_\mp,\quad
(\bgS_{\pm}\,\der\slashed\,\gD_{\mp})^{\,\dag} = 
\bgS_{\mp}\,\der\slashed\,\gD_{\pm}\nonumber
\\
[2mm]
\end{eqnarray}

\end{document}